# Confidence interval for the AUC of SROC curve and some related methods using bootstrap for meta-analysis of diagnostic accuracy studies

Hisashi Noma, PhD[*]
Department of Data Science, The Institute of Statistical Mathematics, Tokyo, Japan
ORCID: http://orcid.org/0000-0002-2520-9949

Yuki Matsushima
Department of Statistical Science, School of Multidisciplinary Sciences, The Graduate University for Advanced Studies, Tokyo, Japan
Department of Biometrics, Otsuka Pharmaceutical Co Ltd, Tokyo, Japan

Ryota Ishii
Biostatistics Unit, Clinical and Translational Research Center, Keio University Hospital, Tokyo, Japan

*Corresponding author: Hisashi Noma
 Department of Data Science, The Institute of Statistical Mathematics
 10-3 Midori-cho, Tachikawa, Tokyo 190-8562, Japan
 TEL: +81-50-5533-8440
 e-mail: noma@ism.ac.jp

**Abstract**

The area under the curve (AUC) of summary receiver operating characteristic (SROC) curve is a primary statistical outcome for meta-analysis of diagnostic test accuracy studies (DTA). However, its confidence interval has not been reported in most of DTA meta-analyses, because no certain methods and statistical packages have been provided. In this article, we provide a bootstrap algorithm for computing the confidence interval of the AUC. Also, using the bootstrap framework, we can conduct a bootstrap test for assessing significance of the difference of AUCs for multiple diagnostic tests. In addition, we provide an influence diagnostic method based on the AUC by leave-one-study-out analyses. We present illustrative examples using two DTA met-analyses for diagnostic tests of cervical cancer and asthma. We also developed an easy-to-handle R package **dmetatools** for these computations. The various quantitative evidence provided by these methods certainly supports the interpretations and precise evaluations of statistical evidence of DTA meta-analyses.

# 1. Introduction

In clinical epidemiology and health technology assessments, meta-analysis for diagnostic test accuracy (DTA) studies has been a standard method for systematic review (Deeks, 2001; Leeflang et al., 2008). The bivariate meta-analysis model has been widely used for these evidence synthesis analyses, as it enables synthesis of the two primary correlated outcomes of diagnostic studies, sensitivity and false positive rate (FPR; = 1−specificity), thereby incorporating their correlation (Reitsma et al., 2005). Also, through the bivariate modeling framework, a corresponding summary receiver operating characteristic (SROC) curve is uniquely identified and it has been used as a primary statistical outcome (Reitsma et al., 2005; Rutter and Gatsonis, 2001).

In most of published DTA meta-analyses, the area under the curve (AUC) of SROC curve is also reported as a primary diagnostic accuracy measure (Deeks, Bossuyt and Gatsonis, 2013; Leeflang et al., 2008; Rutter and Gatsonis, 2001), which corresponds to the $C$-statistic in diagnostic and prognostic studies (Harrell, Lee and Mark, 1996; Steyerberg, 2019). For these prediction accuracy measure, the evaluation of statistical precision is a relevant issue to assess uncertainty of the summary measures, and confidence intervals of the summary sensitivity and FPR are usually presented in practices. However, the confidence interval of the AUC is not reported in most of published DTA meta-analyses. The primary reason is that no closed-form formulae of the confidence interval of AUC has not been provided in the methodological researches and is currently not available in standard software packages. The AUC is one of the most widely used discriminant measures for DTA meta-analyses (Leeflang et al., 2008; Reitsma et al., 2005; Rutter and Gatsonis, 2001), so its uncertainty information would be essential if the AUC is reported in these systematic reviews.

In this article, we present a simple parametric bootstrap algorithm to calculate 95%

confidence interval of the AUC for SROC curve, and provide a R package **dmetatools** (https://github.com/nomahi/dmetatools) that involves an easy-to-handle function to implement the bootstrap algorithm. Also, using the parametric bootstrap algorithm, we can conduct a bootstrap test for assessing significance of the difference of AUCs of multiple diagnostic tests. The R package **dmetatools** also involves an easy-to-handle function to calculate the bootstrap p-value of the test of difference of AUCs.

In addition, recent methodological studies have revealed the relevance of assessments of outliers and their influences in DTA meta-analyses (Matsushima et al., 2020; Negeri and Beyene, 2019). For these influence diagnostic methods, Matsushima et al. (2020) proposed an influential measure based on the AUC, the difference of the AUCs of SROC curves for all population and a leave-one-study-out population (ΔAUC). Since Matsushima et al. (2020)'s method was discussed in the Bayesian framework, the statistical uncertainty of ΔAUC cannot be quantitatively evaluated. However, applying the bootstrap framework, we can obtain a bootstrap distribution of ΔAUC, and can assess its uncertainty quantitatively; the threshold for classifying potentially influential study is clearly determined. This method can also be easily implemented by the R package **dmetatools**.

Also, we illustrate the effectiveness of these methods via applications to DTA meta-analyses of radiological evaluations of lymph node metastases in patients with cervical cancer (Scheidler et al., 1997) and a minimally invasive marker for airway eosinophilia in asthma (Korevaar et al., 2015). Example R codes are also provided in Appendix.

## 2. Bivariate random-effects model and SROC curve for DTA meta-analysis

We consider the Reitsma et al. (2005)'s bivariate random-effects model for sensitivity and FPR of DTA meta-analyses. Let $TP_i$, $FP_i$, $FN_i$ and $TN_i$ be the counts of true positive, false positive, false negative, and true negative participants in the $i$th study, respectively ($i = 1, 2, \ldots, N$). Also, we denote the total numbers of positive and negative diagnoses in the $i$th study as $n_{Ai} = TP_i + FN_i, n_{Bi} = FP_i + TN_i$. At first, we consider the binomial probability models,

$$TP_i \sim Bin(n_{Ai}, p_{Ai}), \ FP_i \sim Bin(n_{Bi}, p_{Bi}).$$

where $p_{Ai}$ and $p_{Bi}$ are the sensitivity and FPR in the $i$th study, respectively. Then, we define the logit-transformed sensitivity and FPR estimators as $y_{Ai} = \text{logit}(TP_i/n_{Ai})$ and $y_{Bi} = \text{logit}(FP_i/n_{Bi})$, and consider the bivariate normal distribution model for $y_{Ai}$ and $y_{Bi}$,

$$\boldsymbol{y}_i \sim N(\boldsymbol{\theta}_i, \boldsymbol{S_i}), \ \boldsymbol{S}_i = \begin{pmatrix} s_{Ai}^2 & 0 \\ 0 & s_{Bi}^2 \end{pmatrix},$$

where $\boldsymbol{y}_i = (y_{Ai}, y_{Bi})^T$ and $\boldsymbol{\theta_i} = (\theta_{Ai}, \theta_{Bi})^T$. $\theta_{Ai}$ and $\theta_{Bi}$ are the logit-transformed sensitivity and FPR for the $i$th study, $\theta_{Ai} = \text{logit}(p_{Ai})$ and $\theta_{Bi} = \text{logit}(p_{Bi})$. Also, $s_{Ai}^2$ and $s_{Bi}^2$ are the variances of $y_{Ai}$ and $y_{Bi}$, which the estimates are obtained as $(TP_i \times FN_i/n_{Ai})^{-1}$ and $(FP_i \times TN_i/n_{Bi})^{-1}$. Note that $TP_i$ and $FP_i$ are assumed to be conditionally independent given the random effect parameters $\theta_{Ai}$ and $\theta_{Bi}$. Then, we consider the random-effects model for the logit-transformed binomial probability parameters,

$$\boldsymbol{\theta}_i \sim N(\boldsymbol{\mu}, \boldsymbol{\Sigma}), \ \boldsymbol{\Sigma} = \begin{pmatrix} \sigma_A^2 & \rho\sigma_A\sigma_B \\ \rho\sigma_A\sigma_B & \sigma_B^2 \end{pmatrix},$$

where $\boldsymbol{\mu} = (\mu_A, \mu_B)^T$, which is the summary logit-transformed sensitivity and FPR. $\sigma_A^2$ and $\sigma_B^2$ correspond to the heterogeneity variances of $\theta_{Ai}$ and $\theta_{Bi}$, and $\rho$ is their correlation coefficient. The bivariate normal-normal model can be handled within the multivariate meta-analysis framework, and the restricted maximum likelihood (REML) estimate of $\{\widehat{\boldsymbol{\mu}}, \widehat{\boldsymbol{\Sigma}}\}$ and its Wald-type confidence interval can be computed by standard multivariate meta-analysis framework (Jackson, Riley and White, 2011; Mavridis and Salanti, 2013; Reitsma et al., 2005). Through the inverse transformation of the REML estimate by $\text{expit}(\mu_A)$ and $\text{expit}(\mu_B)$, the summary sensitivity and FPR estimates are obtained; where $\text{expit}(x) = e^x/\{1 + e^x\}$. Also, the estimated bivariate normal distribution model corresponds to a unique SROC curve by the Rutter and Gatsonis (2001)'s hierarchical regression framework, i.e., a unique SROC curve estimate is identified by the estimated bivariate normal-normal model (Harbord et al., 2007). In practices, the estimated SROC curve and its AUC are reported as summary diagnostic accuracy measures.

## 3. Confidence interval for the AUC of SROC curve

The confidence interval formulae of the AUC for SROC curve like the DeLong's formulae (DeLong, DeLong and Clarke-Pearson, 1988) would be difficult to derive because the AUC estimator is not expressed by closed-form. However, under current computational environment, it is not so problematic. There are various effective Monte Carlo approaches to assess the statistical uncertainty (Hastie, Tibshirani and Friedman, 2009). The representative one is bootstrap method (Davison and Hinkley, 1997; Efron and Tibshirani, 1994). Bootstrap has been a common method for diagnostic and prognostic studies, e.g., in the optimism corrections for internal validation, the bootstrap-based bias correction methods (Harrell et al., 1996; Steyerberg, 2019) are widely applied (Collins et al., 2015; Moons et al., 2015). For the AUC of SROC curve, although the closed-form confidence interval formulae might not be obtained, the bootstrap approach is also effective to calculate the confidence interval. The bootstrap algorithm is as follows.

*Algorithm 1 (Parametric bootstrap for calculating confidence interval of the AUC).*

1. For the bivariate random-effects model, compute the REML estimate $\{\widehat{\boldsymbol{\mu}}, \widehat{\boldsymbol{\Sigma}}\}$.

2. Resample $\boldsymbol{\theta}_1^{(b)}, \boldsymbol{\theta}_2^{(b)}, ..., \boldsymbol{\theta}_N^{(b)}$ from the estimated random-effects distribution $N(\widehat{\boldsymbol{\mu}}, \widehat{\boldsymbol{\Sigma}})$ via parametric bootstrap, $B$ times ($b = 1, 2, ..., B$).

3. Resample $\boldsymbol{y}_1^{(b)}, \boldsymbol{y}_2^{(b)}, ..., \boldsymbol{y}_N^{(b)}$ from the bootstrap sample distribution $N(\boldsymbol{\theta}_i^{(b)}, \boldsymbol{S}_i)$ via parametric bootstrap ($b = 1, 2, ..., B$).

4. Compute the AUC estimates $\widehat{AUC}^{(b)}$ from the $B$ bootstrap samples $\boldsymbol{y}_1^{(b)}, \boldsymbol{y}_2^{(b)}, ..., \boldsymbol{y}_N^{(b)}$ ($b = 1, 2, ..., B$).

5. We can obtain the bootstrap estimate of the sampling distribution of the $\widehat{AUC}$ by the empirical distribution of $\widehat{AUC}^{(1)}, ..., \widehat{AUC}^{(B)}$.

6. The bootstrap confidence interval of the AUC is obtained by the percentiles of the bootstrap distribution; e.g., the 95% confidence interval is obtained by 2.5th and 97.5th percentiles of the bootstrap distribution.

The resultant bootstrap confidence interval is justified by the large sample theory of bootstrap inferences (Davison and Hinkley, 1997; Efron and Tibshirani, 1994). To control the Monte Carlo error of the bootstrap inference, the number of bootstrap resampling should be sufficiently large, usually mentioned to be larger than 1000 (Efron and Tibshirani, 1994). The computation can be easily implemented by the R package **dmetatools**; the example code is provided in Appendix. The computation can usually be completed within a few minutes under the current computational environments.

Besides, some statistical packages for DTA meta-analyses are designed that the arguments are count data of individual studies ($TP_i$, $FP_i$, $FN_i$ and $TN_i$), e.g., **mada** (Doebler, 2019) in R. Then, the parametric bootstrap of Algorithm 1 is inconvenient because the processes 3 and 4 assume the arguments are summary outcome measures $y_1, \ldots, y_N$. As an alternative approach, the processes 3 and 4 can be substituted to a resampling from a binomial probability model as follows.

3'. Resample $TP_i^{(b)}$, and $FP_i^{(b)}$ from the bootstrap sample distribution $Bin\left(n_{Ai}, p_{Ai}^{(b)}\right)$ and $Bin\left(n_{Bi}, p_{Bi}^{(b)}\right)$, respectively; where $p_{Ai}^{(b)} = \text{expit}\left(\theta_{Ai}^{(b)}\right), p_{Bi}^{(b)} = \text{expit}(\theta_{Bi}^{(b)})$.

4'. Compute the AUC estimates $\widehat{AUC}^{(b)}$ from the $B$ bootstrap samples $TP_1^{(b)}, \ldots, TP_N^{(b)}, FP_1^{(b)}, \ldots, FP_N^{(b)}$ $(b = 1, 2, \ldots, B)$.

Thus, even if the bootstrap steps incorporate existing statistical software packages that is designed the arguments to be binomial count data, the bootstrap algorithm can be approximately conducted by substituting the 3 and 4 steps. Both algorithms provide nearly identical results.

## 4. Significance test for the difference of AUCs between two diagnostic methods

In DTA meta-analyses, the comparisons of the summary diagnostic accuracy measures among multiple diagnostic methods are also primary subjects of interests. Reitsma et al. (2005) provided significance tests comparing the summary sensitivity and FPR among multiple diagnostic methods using Wald-type large sample approximations. Besides, there are no closed-form testing methods for comparing AUCs of SROC curves, because the large sample distributions of the AUC estimators are difficult to assess by analytical methods. However, using the bootstrap framework, we can also easily implement a significance test for comparing AUCs among multiple diagnostic methods.

As notations, we suppose outcome measures $y_{1,1}, \ldots, y_{1,N_1}$ and $y_{2,1}, \ldots, y_{2,N_2}$ for two diagnostic tests from $N_1$ and $N_2$ studies. Then, we assume these outcomes follow the bivariate random-effects model respectively, $y_{1,i} \sim N\left(\theta_{1,i}, S_{1,i}\right), \theta_{1,i} \sim N(\mu_1, \Sigma_1)$ and $\boldsymbol{y}_{2,j} \sim N(\boldsymbol{\theta}_{2,j}, \boldsymbol{S}_{2,j})$, $\boldsymbol{\theta}_{2,j} \sim N(\boldsymbol{\mu}_2, \boldsymbol{\Sigma}_2)$ $(i = 1, \dots, N_1; j = 1, \dots, N_2)$. We denote the AUCs of these SROC curves as $AUC_1$ and $AUC_2$. Then, the quantity of interests is the difference of the AUCs,

$$dAUC = AUC_1 - AUC_2$$

and the testing problem here is formulated as

$$H_0: \ dAUC = 0 \ \ \text{vs.} \ H_1: \ dAUC \neq 0$$

The bootstrap p-value of this significance test and the confidence interval of $dAUC$ are calculated by the following bootstrap algorithm.

*Algorithm 2 (Bootstrap test for comparing AUCs between two diagnostic methods).*

1. For the bivariate random-effects models of the two diagnostic tests, compute the REML estimates $\{\widehat{\boldsymbol{\mu}}_1, \widehat{\boldsymbol{\Sigma}}_1\}$ and $\{\widehat{\boldsymbol{\mu}}_2, \widehat{\boldsymbol{\Sigma}}_2\}$. Also, compute the AUC estimates for the two diagnostic tests, $\widehat{AUC}_1$ and $\widehat{AUC}_2$, and their difference $\widehat{dAUC} = \widehat{AUC}_1 - \widehat{AUC}_2$.

2. Resample $\boldsymbol{\theta}_{1,1}^{(b)}, \boldsymbol{\theta}_{1,2}^{(b)}, \dots, \boldsymbol{\theta}_{1,N_1}^{(b)}$ and $\boldsymbol{\theta}_{2,1}^{(b)}, \boldsymbol{\theta}_{2,2}^{(b)}, \dots, \boldsymbol{\theta}_{2,N_1}^{(b)}$ from the estimated random-effects distributions $N(\widehat{\boldsymbol{\mu}}_1, \widehat{\boldsymbol{\Sigma}}_1)$ and $N(\widehat{\boldsymbol{\mu}}_2, \widehat{\boldsymbol{\Sigma}}_2)$, respectively, $B$ times $(b = 1,2,\dots,B)$.

3. Resample $\boldsymbol{y}_{1,1}^{(b)}, \boldsymbol{y}_{1,2}^{(b)}, \dots, \boldsymbol{y}_{1,N_1}^{(b)}$ and $\boldsymbol{y}_{2,1}^{(b)}, \boldsymbol{y}_{2,2}^{(b)}, \dots, \boldsymbol{y}_{2,N_2}^{(b)}$ from the bootstrap sample distribution $N(\boldsymbol{\theta}_{1,i}^{(b)}, \boldsymbol{S}_{1,i})$ and $N(\boldsymbol{\theta}_{2,j}^{(b)}, \boldsymbol{S}_{2,j})$ $(i = 1, \dots, N_1; j = 1, \dots, N_2; b = 1,2,\dots,B)$.

4. Compute the AUC estimates $\widehat{AUC}_1^{(b)}$ and $\widehat{AUC}_2^{(b)}$ from the $B$ bootstrap samples $\boldsymbol{y}_{1,1}^{(b)}, \boldsymbol{y}_{1,2}^{(b)}, \dots, \boldsymbol{y}_{1,N_1}^{(b)}$ and $\boldsymbol{y}_{2,1}^{(b)}, \boldsymbol{y}_{2,2}^{(b)}, \dots, \boldsymbol{y}_{2,N_2}^{(b)}$ $(b = 1,2,\dots,B)$.

5. Compute the difference of the AUCs, $\widehat{dAUC}^{(b)} = \widehat{AUC}_1^{(b)} - \widehat{AUC}_2^{(b)}$. Then, we can obtain the bootstrap estimate of the sampling distribution of the $\widehat{dAUC}$ by the empirical distribution of $\widehat{dAUC}^{(1)}, \dots, \widehat{dAUC}^{(B)}$.

6. The bootstrap p-value is obtained by using the bootstrap distribution of $\widehat{dAUC}$ as the reference distribution. Also, the confidence interval of the $dAUC$ is obtained by the percentiles of the bootstrap distribution; e.g., the 95% confidence interval is obtained by 2.5th and 97.5th percentiles of the bootstrap distribution.

The resultant bootstrap p-value and confidence interval is justified by the large sample theory of bootstrap inferences (Davison and Hinkley, 1997; Efron and Tibshirani, 1994). The computation can also be easily implemented by the R package **dmetatools**, and can usually be completed within a few minutes under current computational environments. If the analysists would like to incorporate existing statistical software packages for the computation of AUC that is designed the arguments to be the binomial count data, the bootstrap processes can be substituted to the 3' and 4' steps mentioned in the previous section.

## 5. Influence diagnostics by AUC

Outlying studies that have extreme profiles in the overall populations can seriously influence the results and conclusions of meta-analysis, so the influence diagnostics are addressed as relevant methods for meta-analyses (Hedges and Olkins, 1985; Viechtbauer and Cheung, 2010). Negeri and Beyene (2019) and Matsushima et al. (2020) generalized the influence diagnostic methods of univariate meta-analysis, and proposed effective methods for DTA meta-analysis. Especially, Matsushima et al. (2020) proposed a new diagnostic measure based on the change of AUC by leave-one-study-out analysis. For the notations, let $\widehat{\text{AUC}}$ be the AUC estimate from the all $N$ studies, and $\widehat{\text{AUC}}^{(-i)}$ be the AUC estimate from $N-1$ studies that excludes $i$th study. Then, the AUC-based influential diagnostic statistic (Matsushima et al., 2020) is given as

$$\Delta\widehat{\text{AUC}}(i) = \widehat{\text{AUC}} - \widehat{\text{AUC}}^{(-i)}$$

The influential measure $\Delta\widehat{\text{AUC}}^{(-i)}$ quantitatively assesses the impact of $i$th study for the AUC estimate ($i = 1, 2, \ldots, N$). Although other influential measures based on summary sensitivity, FPR, and diagnostic odds-ratio (Matsushima et al., 2020; Negeri and Beyene, 2019) were also proposed, the AUC is another primary diagnostic accuracy measure, it would be a useful and intuitively interpretable measure to assess the influences of individual studies by the changes of AUC. However, Matsushima et al. (2020) discussed their methods within the Bayesian methodological framework, the threshold determination is possibly a difficult problem in practices. In the frequentist formulation, we can quantitatively assess the uncertainty of $\widehat{\Delta AUC}(i)$ if the sampling distribution is accurately estimated, and can select the threshold reasonably. Although the sampling distribution cannot also be estimated analytically, we can apply the bootstrap schemes of the previous sections for the estimation. The bootstrap algorithm is provided as follows.

*Algorithm 3 (Bootstrap for estimating the sampling distribution of $\widehat{\Delta AUC}(i)$).*

1. For the bivariate random-effects model, compute the REML estimate $\{\hat{\boldsymbol{\mu}}, \hat{\boldsymbol{\Sigma}}\}$.

2. Resample $\boldsymbol{\theta}_1^{(b)}, \boldsymbol{\theta}_2^{(b)}, \dots, \boldsymbol{\theta}_N^{(b)}$ from the estimated random-effects distribution $N(\hat{\boldsymbol{\mu}}, \hat{\boldsymbol{\Sigma}})$ via parametric bootstrap, $B$ times ($b = 1, 2, \dots, B$).

3. Resample $\boldsymbol{y}_1^{(b)}, \boldsymbol{y}_2^{(b)}, \dots, \boldsymbol{y}_N^{(b)}$ from the bootstrap sample distribution $N(\boldsymbol{\theta}_i^{(b)}, \boldsymbol{S}_i)$ via parametric bootstrap ($b = 1, 2, \dots, B$).

4. Compute the AUC-based influential diagnostic statistic $\widehat{\Delta AUC}(i)^{(b)}$ from the $B$ bootstrap samples $\boldsymbol{y}_1^{(b)}, \boldsymbol{y}_2^{(b)}, \dots, \boldsymbol{y}_N^{(b)}$ ($b = 1, 2, \dots, B$).

5. We can obtain the bootstrap estimate of the sampling distribution of the $\widehat{\Delta AUC}(i)$ by the empirical distribution of $\widehat{\Delta AUC}(i)^{(1)}, \dots, \widehat{\Delta AUC}(i)^{(B)}$.

6. The threshold can be determined by the percentiles of the bootstrap distribution; e.g., 2.5th and 97.5th percentiles of the bootstrap distribution are adopted.

Based on the bootstrap method, we can assess how the corresponding study has influence in the overall population. The computation can also be implemented by the R package **dmetatools**, and an example code is provided in Appendix.

## 6. Applications

### *6.1 Confidence interval for the AUC of SROC curve*

As illustrative examples, we present the summary sensitivity and FPR estimates for DTA meta-analyses of radiological evaluations of lymph node metastases in patients with cervical cancer by 44 studies of Scheidler et al. (1997) in Table 1. This DTA meta-analysis evaluated three imaging techniques for the diagnosis of lymph node metastasis in women with cervical cancer, lymphangiography (LAG) based on the presence of nodal-filling defects (N=17), computed tomography (CT; N=17), and magnetic resonance imaging (MRI) for nodal enlargement (N=10). The summary sensitivity and FPR with the 95% confidence intervals were estimated by the REML method using the bivariate random-effects model. In Table 1, the heterogeneity standard deviation estimates $\sigma_A, \sigma_B$ are also provided. The confidence intervals are usually reported because they can be computed by closed-form formulae. Besides, the SROC curves for the three imaging techniques estimated by the above method are presented in Figure 1. They provide graphical summaries for the diagnostic accuracies of these imaging techniques, and can be interpreted similarly with the common ROC curve for ordinary diagnostic and prognostic studies in clinical epidemiology. The point estimates of the AUCs are easily calculated by standard software packages (e.g., **mada** (Doebler, 2019) in R), but the closed-form formulae of confidence interval has not been developed currently. Effective computational methods for the confidence interval and their computational package would be needed for practitioners of these meta-analyses.

We applied the bootstrap algorithm to the DTA meta-analysis of cervical cancer. We provided the resultant 95% confidence intervals of the AUCs for the three imaging techniques (LAG, CT, and MRI) in the third row of Table 1. The numbers of bootstrap resampling were set to 2000, consistently. Although the AUC estimates of CT and MRI were certainly larger than that of LAG, the 95% confidence intervals of the AUCs indicate there were certain statistical uncertainties for these estimates. These confidence intervals would provide relevant information for interpretations of the diagnostic accuracies.

### 6.2 Comparing the AUCs of multiple SROC curves

For the cervical cancer dataset, we also applied the bootstrap methods for comparing the AUCs of multiple SROC curves. We provided the resultant bootstrap p-values and 95% confidence intervals of the dAUCs for pairwise comparisons of the three imaging techniques (CT vs. LAG, MRI vs. LAG, and MRI vs. CT) in Table 2. All of the p-values were larger than 0.05 and the AUCs were not significantly different. However, the 95% confidence intervals of dAUC indicate the AUCs of CT and MRI might certainly larger than that of LAG. Note that Reitsma et al. (2005) conducted significance tests among the three imaging techniques for the differences of summary sensitivity, specificity (= $1-$FPR), and diagnostic odds-ratio. They reported some of these measures were significantly different for CT vs. LAG and MRI vs. LAG. Although the results of significance tests were inconsistent, it is common because these diagnostic measures aim to assess different characteristics. However, the AUC has been a common primary diagnostic accuracy measure, so they would provide additional useful information for more precise technology assessments. The numbers of bootstrap resampling were also set to 2000, consistently.

### 6.3 Influence diagnostics by AUC

For illustrations of the AUC-based influential diagnostic method, we applied the proposed method to a meta-analysis of diagnostic accuracy of airway eosinophilia in asthma by Korevaar et al. (2015). Korevaar et al. (2015) conducted DTA meta-analyses of minimally invasive markers for diagnostic of eosinophilic airway inflammation, and we consider their meta-analysis for the fraction of exhaled nitric oxide here ($N$ = 12). The SROC curve is presented at Figure 2. The AUC estimate was 0.757 (95%CI: 0.698, 0.791). The influential statistic $\widehat{\Delta \text{AUC}}(i)$ assesses how the AUC changes when $i$th study is deleted. The numbers of bootstrap resampling were set to 2000.

The results of the influence diagnostics are presented in Table 3. The AUC estimates by leave-one-out analyses varied on 0.729-0.780. The 5th and 8th studies were the most influential ones for changing the AUC estimate. The SROC curves obtained by leave-one-out analyses by the two studies are shown in Figure 3. Leaving the 5th study, the SROC estimate shifted downward, because this study had relatively large sensitivity estimate. The AUC estimate was decreased to 0.729, and $\widehat{\Delta \text{AUC}}(5) = -0.028$. Also, since the 8th study had relatively small sensitivity and large FPR estimates, deleting this study, the SROC estimate shifted upward and the AUC estimate was increased to 0.780, i.e., $\widehat{\Delta \text{AUC}}(8) = 0.023$. For both studies, the realized values of the influential statistic exceeded the 95% bootstrap interval. Thus, they would be interpreted to have influential profiles that exceeded the reasonable ranges explained by the random variations. Thus, these studies should be interpreted as influential, and be carefully re-investigated for their background information that possibly influence the interpretations of the overall results. Note that outlying studies and influential studies can be different, and this method focuses to detect the latter. The detected studies are usually influential ones to the AUC estimate, but are not outliers.

## 7. Discussion

In this paper, we provided a bootstrap algorithm to compute a confidence interval of the AUC of SROC curve and easy-to-handle software package for DTA meta-analysis. SROC curve has been already a primary statistical outcome of DTA meta-analysis (Deeks et al., 2013; Leeflang et al., 2008; Rutter and Gatsonis, 2001) and is currently adopted in most published papers. The statistical uncertainty information is also essential and it should be reported as a standard outcome of these studies. As an alternative approach, the large sample distribution of the AUC might be theoretically derived in future studies. Then, another confidence interval using a closed-form formulae is possibly available, but currently it is uncertain. However, if such formulae is developed, the bootstrap approach is also effective, because its large sample approximation can have better coverage performances compared with Wald-type confidence intervals (Noma et al., 2018; Ukyo et al., 2019), and the computational cost is not problematic under modern computational environments.

In addition, the comparisons of multiple diagnostic tests are important subjects in DTA meta-analysis (Reitsma et al., 2005). As like the confidence interval of AUC, the closed-form formulae of the standard error estimate and test statistic of dAUC would be difficult to derive theoretically. Thus, the bootstrap approach would be a feasible and effective approach currently. Its computational cost is also not problematic, so it is a tractable computational solution for practices.

For the influence diagnostics based on the AUC, the bootstrap approach enables quantitative evaluation of the realized value of ΔAUC and can provide reasonable thresholds to assess influential study (Negeri and Beyene, 2019; Noma et al., 2020). Note that the ΔAUC is not a standardized measure like the studentized residual for linear regression analysis (Belsley, Kuh and Welsch, 1980), the absolute values are not comparable among the $N$ studies. However, the bootstrap distributions provide clear information concerning the relative sizes of them by their sampling distributions. The bootstrap approach would provide quantitative summaries of them for practices of DTA meta-analysis.

The bootstrap methods provided in this article can be easily implemented by installing the R package **dmetatools** (https://github.com/nomahi/dmetatools). The analyses can be conducted by simple codes and are easy to handle for non-statisticians. These statistical outcomes would be recommended to involve in DTA meta-analyses if the authors report the AUC of SROC curve. These various quantitative evidence certainly supports the interpretations and precise evaluations of statistical evidence for DTA meta-analyses.

## Acknowledgements

This study was supported by Grant-in-Aid for Scientific Research from the Japan Society for the Promotion of Science (Grant number: JP19H04074).

**Appendix: R codes for implementation of the bootstrap methods**

```
# Installation of the dmetatools package
require(devtools)
devtools::install_github("nomahi/dmetatools")

# Computation of the 95% CI for the AUC of SROC curve
library(dmetatools)
data(cervical)
CT <- cervical[cervical$method==1,]
LAG <- cervical[cervical$method==2,]
MRI <- cervical[cervical$method==3,]

AUC_boot(CT$TP,CT$FP,CT$FN,CT$TN)
AUC_boot(LAG$TP,LAG$FP,LAG$FN,LAG$TN)
AUC_boot(MRI$TP,MRI$FP,MRI$FN,MRI$TN)

# Comparing the differences of AUCs
AUC_comparison(CT$TP,CT$FP,CT$FN,CT$TN,LAG$TP,LAG$FP,LAG$FN
,LAG$TN)
AUC_comparison(MRI$TP,MRI$FP,MRI$FN,MRI$TN,LAG$TP,LAG$FP,LA
G$FN,LAG$TN)
AUC_comparison(MRI$TP,MRI$FP,MRI$FN,MRI$TN,CT$TP,CT$FP,CT$F
N,CT$TN)

# Influence diagnostics by the delta-AUC
data(asthma)
AUC_IF(asthma$TP, asthma$FP, asthma$FN, asthma$TN)
```

For more details, please see the help files of the **devtools** package and the web page (https://github.com/nomahi/dmetatools).

**Table 1.** Estimates and 95% confidence intervals for the summary sensitivity, FPR, and AUC of the SROC curve for the cervical cancer data.

|  | Sensitivity | FPR | AUC |
|---|---|---|---|
| CT | 0.489 (0.372, 0.608) | 0.088 (0.062, 0.123) | 0.864 (0.659, 0.891) |
| Between-studies SD[†] | 0.774 | 0.519 | |
| LAG | 0.658 (0.590, 0.719) | 0.190 (0.135, 0.261) | 0.755 (0.617, 0.809) |
| Between-studies SD[†] | 0.320 | 0.724 | |
| MRI | 0.556 (0.380, 0.718) | 0.063 (0.034, 0.112) | 0.893 (0.646, 0.932) |
| Between-studies SD[†] | 0.939 | 0.802 | |

[†] Between-studies standard deviation (SD) estimates for logit-transformed outcomes

**Table 2.** Estimate and 95% confidence interval for the dAUC for the cervical cancer data. Also, p-value for the test of "dAUC=0" is presented.

|            | dAUC                    | p-value |
|------------|-------------------------|---------|
| CT vs. LAG | 0.110 (−0.062, 0.228)   | 0.130   |
| MRI vs. LAG| 0.139 (−0.097, 0.271)   | 0.179   |
| MRI vs. CT | 0.029 (−0.190, 0.181)   | 0.668   |

**Table 3.** Results of the influential diagnostics based on AUC for the asthma data.

| | AUC | ΔAUC | Bootstrap 2.5th percentile | Bootstrap 97.5th percentile |
|---|---|---|---|---|
| All studies | 0.757 | | | |
| Excluding 5th study | 0.729 | −0.028 | −0.026 | 0.024 |
| Excluding 8th study | 0.780 | 0.023 | −0.024 | 0.021 |
| Excluding 1st study | 0.768 | 0.011 | −0.011 | 0.014 |
| Excluding 4th study | 0.762 | 0.005 | −0.021 | 0.017 |
| Excluding 6th study | 0.753 | −0.004 | −0.010 | 0.013 |
| Excluding 12th study | 0.753 | −0.004 | −0.034 | 0.026 |
| Excluding 2nd study | 0.754 | −0.003 | −0.011 | 0.015 |
| Excluding 11th study | 0.760 | 0.003 | −0.022 | 0.019 |
| Excluding 3th study | 0.755 | −0.003 | −0.007 | 0.013 |
| Excluding 9th study | 0.759 | 0.001 | −0.015 | 0.012 |
| Excluding 10th study | 0.756 | −0.001 | −0.012 | 0.012 |
| Excluding 7th study | 0.757 | 0.000 | −0.015 | 0.012 |

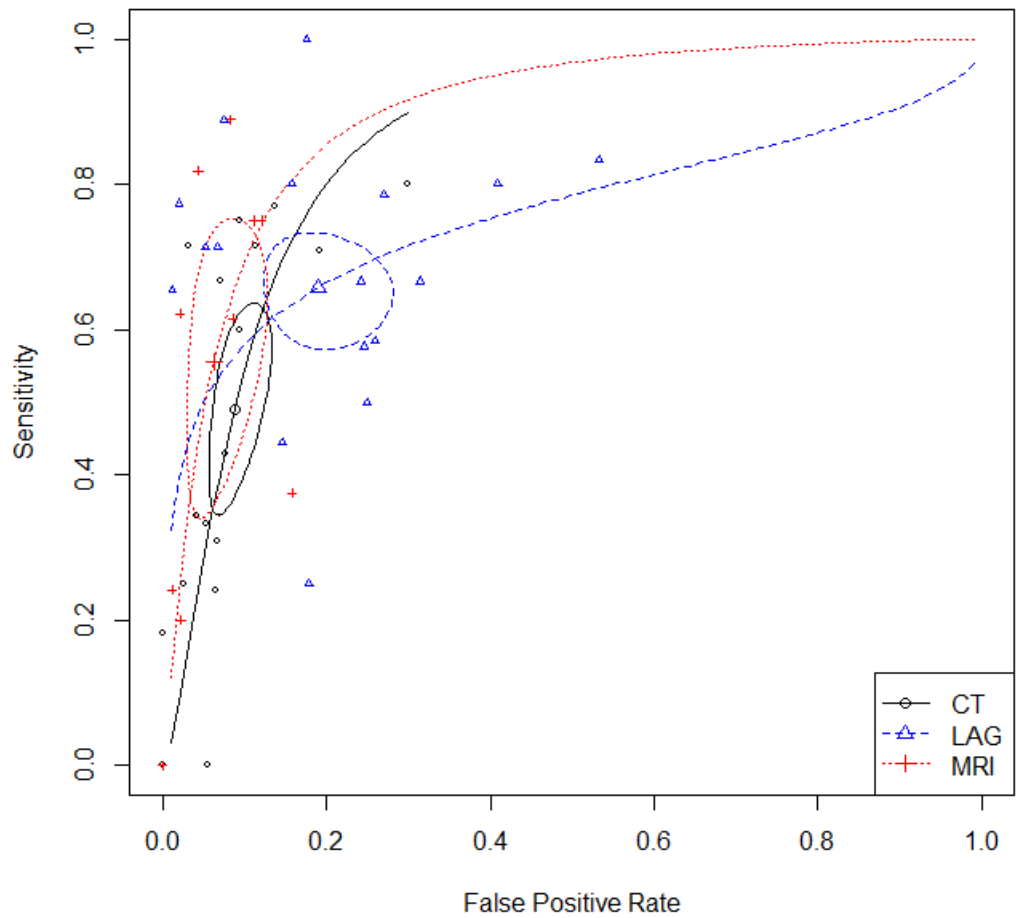

**Figure 1.** SROC curves for the cervical cancer data. The sensitivity and FPR estimates of individual studies, and their summary estimates with 95% confidence regions are also presented.

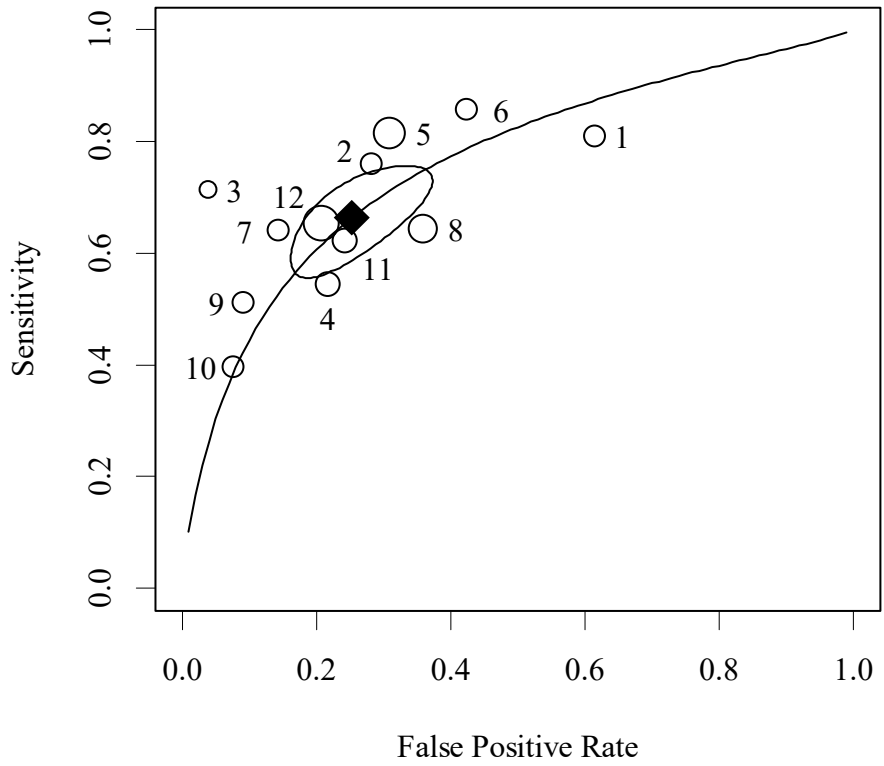

**Figure 2.** SROC curve for the asthma data. The sensitivity and FPR estimates of individual studies, and their summary estimates with 95% confidence regions are also presented.

(a) Leave-one-out analysis excluding the 5th study        (b) Leave-one-out analysis excluding the 8th study

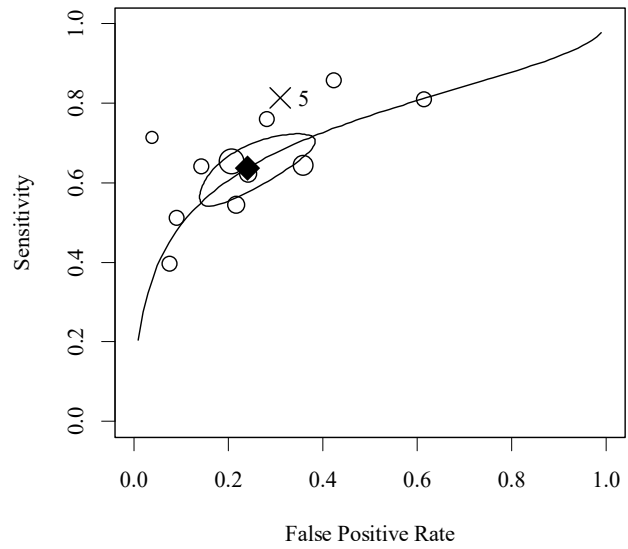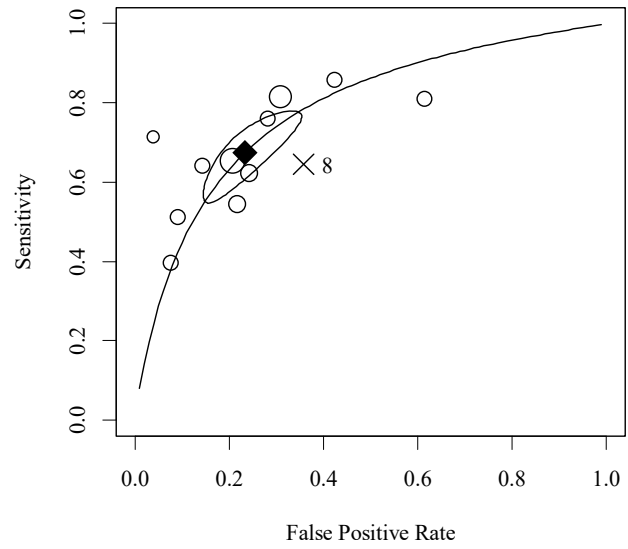

**Figure 3.** SROC curves that deleted the 5th and 8th studies that were detected as potential influential studies for the asthma data.